\documentclass[letterpaper]{article}
\usepackage{aaai}
\usepackage{times}
\usepackage{helvet}
\usepackage{courier}
\usepackage{url}
\begin{document}
%
\title{Be In The Know: Connecting News Articles to Relevant Twitter Conversations}
\author{Bichen Shi, Georgiana Ifrim, Neil Hurley \\\\
Insight Centre for Data Analytics\\
University College Dublin\\
Belfield, Dublin 4, Ireland\\
}
\maketitle
\begin{abstract}
\begin{quote}
In the era of data-driven journalism, data analytics can deliver tools to support journalists in connecting to new and developing news stories, 
e.g., as echoed in micro-blogs such as Twitter, the new citizen-driven media. In this paper, we propose a framework for tracking and automatically 
connecting news articles to Twitter conversations as captured by Twitter hashtags. For example, such a system could alert journalists about news 
that get a lot of Twitter reaction, so that they can  investigate those conversations for new developments in the story, promote their article to a set of
interested consumers, or discover general sentiment towards the story. Mapping articles to appropriate hashtags is nevertheless very challenging, 
due to different language styles used in articles versus tweets, the streaming aspect of news and tweets, as well as the user behavior when marking 
certain tweet-terms as hashtags. As a case-study, we continuously track the RSS feeds of Irish Times news articles and a focused Twitter stream over 
a two months period, and present a system that assigns hashtags to each article, based on its Twitter echo. 
We propose a machine learning approach for classifying and ranking article-hashtag pairs. 
Our empirical study shows that our system delivers high precision for this task.

\end{quote}
\end{abstract}

\section{Introduction}

Since its start in 2006, Twitter has established itself as an alternative media source, with its 500 million users sending more than 500 million tweets daily on every possible topic. The 140 character messages called tweets, are typically grouped around the same subject by dedicated hashtags, e.g., political events: \#btw13 (German Elections); crises: \#Egypt, \#USshutdown; natural disasters:  \#Haiyan (Philipines typhoon);  epidemics: \#H1N1; sports: \#worldcup, \#ashes; celebrities: \#Messi, \#royalbaby. 
Sometimes the news flows from the mainstream media to Twitter, and sometimes it is the Twitter users that first set the ground for breaking news stories, e.g., the 2009 airplane landing on the river Hudson.

Increasingly, Twitter conversations and calls to action that mobilize masses have dedicated hashtags, as showcased by recent world events, e.g., \#ArabSpring, \#Syria, \#freethe7. Twitter hashtags thus lead to the formation of ad hoc publics around specific themes and topics  without the need for the users to be otherwise explicitly connected \cite{bruns2011use}.
In our study we learned, for example, that for popular sporting events involving the national teams in Ireland, Twitter users prefer the hashtag \#coybig (short for Come On You Boys in Green). For a recent political scandal in Germany, related to BMW funding Angela Merkel's political campaign, one of the hashtags preferred by users for commenting on this news was \#buy\_a\_merkel.  Hashtags can convey information about the community that uses them or the sentiment of the messages they group. In the US, for example, the hashtag \#RacismEndedWhen groups tweets that mock a GOP tweet that seemed to suggest that racism had been ended. For an outsider, or even for an insider that doesn't continuously track the massive Twitter activity, it is close to impossible to stay in the know when it comes to the right hashtags or users to follow, for current and developing news stories. Nevertheless for journalists in particular, it is vital to get to the right hashtags quickly, in order to be able to follow new developments on topics of interest. Data analytics techniques can provide  tools that link news stories to the relevant Twitter conversations. 

Automatically mapping news articles to appropriate hashtags (where a hashtag is seen as a group of tweets forming a conversation around it) can be very challenging. This is due to different language styles used in the two types of data (e.g., clean, long articles versus messy, short tweets), the fast paced streaming aspect of both news and tweets (matching two streams moving at different speeds), as well as user behaviour when coining certain tweet-terms as hashtags. To showcase the third issue, in Table \ref{table:example-article} we present an example news article and the categories we identified for the hashtags retrieved for it, in an initial pre-processing stage.
The article is about Irish politics, in particular, the 2013 referendum to adopt a unicameral parliamentary system by abolishing one of the current two houses of parliament, the Seanad. This referendum was proposed by the Irish prime minister, Enda Kenny, and his party Fine Gael (FG).
The hashtags retrieved for this article in an initial pre-processing step, range from highly specific and relevant (e.g., \#seaned, \#irelandsaysno), to general but still relevant (e.g., \#ireland, \#irishpolitics, \#rtept for RTE Prime Time, a TV show broadcasting a debate about the  Seanad referendum), to abusive but potentially relevant (\#caugthrotten referring to Irish PM Enda Kenny's role in this referendum), to irrelevant (e.g., \#tobaccodirective, \#mentalhealth, referring to other topics).
We can see from this example that an approach that can accurately filter irrelevant hashtags and rank relevant hashtags can deliver value by connecting to the right Twitter conversations. 

In this paper we propose a framework for connecting news articles from mainstream media to their echo on the Twitter stream. We discuss the data collection process for continuously gathering, processing and connecting a stream of news articles and a focused Twitter stream relevant to the tracked news stories.
We analyze relevant features and propose a machine learning algorithm for ranking hashtags for a given news story. 
Our experiments show that our system can achieve high precision on this task.
The rest of the paper is organized as follows. Section 2 discusses related work and our contributions. In Section 3 we explain the data collection process, while in Section 4 we describe the process of modeling 
hashtag ranking as a learning problem. In Section 5 we discuss our results and Section 6 concludes with directions for future work.
\begin{table*}[t!]
\centering
\caption{Example news article and initially retrieved hashtags (before learning algorithm is applied).}
\label{table:example-article}
\begin{tabular}{|l|l|l|l|l|}
\hline
News Article								& Retrieved Hashtags (no learning)	& Hashtag Category\\\hline
Headline: \emph{FG fears day of reckoning} 		& \#seanad, \#irelandsaysno, \#enda		& Relevant (Specific) \\
\emph{over Enda Kenny's Seanad gamble}		&							&	\\
										& \#ireland, \#rtept,\#news	& Relevant (General) \\
Sub-headline:  \emph{There is deep concern }		&							& \\
\emph{within the Fine Gael ranks that its }			& \#caughtrotten, \#whip			& Relevant (Abusive) \\
\emph{populist referendum campaign misfired}		& 							& \\
\emph{so badly}							& \#tobaccodirective, \#mentalhealth	& Irrelevant \\\hline
 \end{tabular}
\end{table*}

\section{Related Work}

Recent years have seen an explosion of research work analyzing social media (e.g., most prominently the micro-blog Twitter) and the connection between traditional media and this new form of reporting.
Twitter studies focus on topics such as detecting political leaning from tweets \cite{boutet2012s,himelboim2013birds}, sentiment and opinion mining \cite{pak2010twitter,liu2012survey,luo2012opinion}, summarising sporting, economic and other events using Twitter \cite{popescu2011dancing,nichols2012summarizing,hu2012were}; analysing
news spread through the network \cite{kunegis2011bad,artzi2012predicting}; content curation \cite{agichtein2008finding,castillo2011information};
user influence and authority \cite{bakshy2011everyone,romero2011influence};
detecting breaking news stories from the massive tweet stream, potentially ahead of traditional media reporting \cite{sankaranarayanan2009twitterstand,kwak2010twitter}.

Among the diverse investigations of Twitter data, two categories are most relevant to this paper, namely, research that has focussed on hashtag recommendation or retrieval and studies considering the mapping between news articles and microblogs.  

\noindent{\bf Hashtag Recommendation.} Tag recommendation for tagging systems such as Last.fm and Delicious has been studied in a number of works such as \cite{krestel2009latent} that applies topic modelling using Latent Dirichilet Allocation (LDA) to the problem.  Focusing in particular on hashtag retrieval over a Twitter corpus,   in \cite{efron2010hashtag},  language modelling is used to find hashtags given a keyword query. A model of each hashtag is learned from the set of tweets that contain the tag as a multinomial distribution over terms.  Hashtags are ranked according to the KL divergence of their corresponding model to the query model.  
The related issue of keyphrase extraction from Twitter is studied in \cite{zhao2011topical}. Rather than associate a set of existing hashtags with individual tweets, the goal in this work is to recommend keyphrases  generated from the entire vocabulary.  The method follows three steps.  In the first step, a topic model of the tweets is built, using a modified version of LDA suitable for short documents, proposed in  \cite{zhao2011comparing}, in which  a tweet is generated from just a single topic along with a background noise `topic' learned from the entire corpus.  Next, a pagerank-like algorithm is run over each topic, to identify the most influential terms associated with the topic.  Finally, keyphrases are generated by combining the top ranked terms and are ranked according to their relevance and interestingness.  In \cite{DBLP:conf/coling/DingZH12} the issue of recommending hashtags to untagged tweets is addressed. An LDA topic model is used to categorise tweets into topics and a translation probability maps topics to hashtags.  The method is modified in \cite{ding2013learning}  by replacing standard LDA with the topic model of \cite{zhao2011comparing}.

\noindent{\bf News and Tweets.} Work that investigates the connection between news and Twitter includes \cite{vstajner2013automatic}. Given a set of tweets that specifically mention the URL of a given article, this work focuses on a method to filter this set  into a subset of most interesting tweets. The authors use four indicators of interestingness, namely informativeness, opinionatedness, popularity and authority to filter the initial set. TweetMogaz \cite{Magdy:2013}, a system for microblog search and filtering, aims to find tweets relevant to regional news. It relies on a curated list of \emph{key players} from which to collect an initial set of relevant tweets. The initial set is augmented, by firstly extracting a set of keywords from news sites and searching for tweets containing these keywords. The \emph{keyword} tweets are filtered by training a classifier using the \emph{key player} tweets as positive examples and a set of random tweets as negative examples. The \emph{keyword} tweets that are classified as positive are retained.  
In \cite{DBLP:conf/icwsm/LehmannCLZ13}, a system is proposed to support journalists in rapidly detecting follow-ups  to their news articles that break on Twitter. Here again, the data extraction process starts with a set of tweets that mention the URL of a given news article. From this initial set of tweets, a set of users that tweeted the article within a limited time after its first tweet is gathered, and the tweets of this user group or \emph{crowd} over a following time-period are analysed to discover new tweets related to the story.
Other works investigated automatic news detection from tweets  \cite{subavsic2011peddling}, recommending news articles using tweets \cite{phelan2011terms}, forecasting the popularity of news using Twitter \cite{bandari2012pulse}, or enhancing news articles with information extracted from Twitter, such as \emph{comment tweets} \cite{kothari2013detecting}.

Our work differs from the above research in a number of ways. In particular, we address hashtag recommendation in a streaming context, with a requirement that the model be updated on a daily basis. Rather than apply topic modeling on a large, static Twitter corpus, containing potentially many diverse topics, we attempt to filter irrelevant tweets directly by using the news articles to be hashtagged in order to focus the data collection from the Twitter stream. Nevertheless, unlike other work on connecting articles and microblogs, we avoid seeding our data collection with a curated user group or with tweets that specifically mention the articles in question (via the URL).  As discussed later, our dynamic-keyword Twitter stream allows for a wide set of tweets to be gathered, while ensuring that the collection contains relevant tweets with high probability. We believe that our search strategy provides sufficient breadth to allow high recall in gathering relevant hashtags, while avoiding being drowned in a vast sea of Twitter noise. We alternate this high recall with a high precision oriented step, by using a learning approach to rank the retrieved hashtags for each article.

\noindent{\bf Our contributions} are as follows: (1) we propose a focussed Twitter data collection strategy based on dynamic keyword extraction from news articles; (2) we formulate a learning algorithm for assigning hashtags to news articles; (3) we deliver a system for matching a daily news stream and a relevant Twitter conversation stream.

\section{Data Collection}

All the data collected for this study is available upon request for research purposes.

\subsection{News Articles from RSS Feeds}

We gathered the news articles streamed on the Irish Times RSS feeds between October 7, 2013 and November 30, 2013, by polling the RSS feeds every 5 minutes, yielding a total of 4,862 unique articles. The Irish Times is an Irish mainstream media outlet, that covers Irish news and high impact world news. There are three RSS feeds for general news and more specialised business and sport news. Each article has a headline, a one paragraph description that summarises the article (sub-headline), and the article body. Although there is additional meta-data associated with each article, in the form of manually assigned topics and some named entity annotations, these tend to be scarce and noisy, thus we currently do not use this meta-data in our approach.

We use Python for scrapping the urls from the RSS feed, downloading the html files and processing the text. Table \ref{table:rss-stream-stats} shows statistics on the number of news articles retrieved daily from the Irish Times RSS feed. The minimum number of articles corresponds to a Sunday (October 27, 2013), while the maximum corresponds to the release of the Irish budget for 2014 (October 15, 2013).
To obtain at-a-glance coverage of the newsfeed, needed for collecting a focused Twitter stream (as explained below), we extract representative keywords for each downloaded article. 
We parse the headline and sub-headline, part-of-speech tag this text, and extract nouns and named entities using shallow parsing techniques and heuristics (e.g., we extract Aer Lingus, Enda Kenny, etc.). We do not use the article-body for keyword extraction, since it poses risks of topic drift and noise.
For example, for the news article in Table \ref{table:example-article} we extract the keywords \emph{enda kenny, fine gael, fg, fears, seanad}. 
\begin{table}
\centering
\caption{Statistics on the daily number of articles in the Irish Times RSS stream.}
\label{table:rss-stream-stats}
\begin{tabular}{|c|c|c|c|c|}
\hline
Min. & Median & Mean & Max & Stdev \\\hline
 87 & 168 & 159 & 213 & 32.47\\\hline
 \end{tabular}
\end{table}
 
\subsection{Focused Twitter Stream}

We have experimented with several strategies for collecting Twitter data relevant to the daily news stream. Since we are interested in continuously streaming news and corresponding tweets, we use the Twitter Streaming API\footnote{https://dev.twitter.com/docs/streaming-apis}. The Streaming API can be employed with either  keywords (words or phrases), geographical boundary boxes or user ID. Studies show \cite{morstatter2013sample} that the Twitter Streaming API provides access to 1\% up to 40\% of the public tweet-stream\footnote{http://www.brightplanet.com/2013/06/twitter-firehose-vs-twitter-api-whats-the-difference-and-why-should-you-care}. 
This data may nevertheless be irrelevant to our focused information need. We looked at 4 alternatives to gather Twitter streaming data: using a curated set of users (200 Irish journalists), a static set of keywords (names of cities in Ireland), a geo-focused stream (using location coordinates to capture Irish tweets) and a dynamic set of keywords extracted from the stream of RSS news articles every 30 mins each day.
Comparing the 4 streams, the dynamic-keyword stream was considerably larger, with a total of 23,362,818 
unique tweets, as compared to 291,141
in the curated user stream, 
1,629,678
in the geo-stream, and 8,527,952 
in the static keyword stream. 
In this work we focus on the fourth method: the dynamic-keyword-focused Twitter stream.
Due to the restriction of the Streaming API to use a maximum of 400 keywords, we limit the number of keywords extracted from articles by giving preference to named entities and frequent nouns .
 
Additionally, we noticed that in order to get relevant tweets, it helps if we constrain each tweet returned by the Twitter API to contain at least two article keywords. We achieve this by splitting our original keyword set, into individual keywords, and creating all possible permutation pairs as our final keyword set, with the constraint that we freeze named entities. For example, for the article in Table \ref{table:example-article}, we process the keyword set \emph{enda kenny, fine gael, fg, fears, seanad} by keeping the named entities and permuting the single keywords to form pairs, as shown in Table \ref{table:keywords-permutation}. We apply this process every 30 minutes to \emph{all} the RSS articles downloaded up to that point in time, pool the keywords together, and re-connect with the Streaming API using the updated keyword list.
\begin{table}
\centering
\caption{Processed keyword set by permutation.}
\label{table:keywords-permutation}
\begin{tabular}{|c|c|}
\hline
Original keywords/phrases 	& Final keywords/phrases \\\hline
enda kenny 				& enda kenny\\
fine gael 					& fine gael\\
fg 						& fg fears\\
fears  					& fg seanad\\
seanad  					& fears seanad\\\hline
 \end{tabular}
\end{table}

 Through this process we aim to retrieve a large set of relevant tweets whilst not being restricted to a set of manually curated user lists, locations or keywords. The problem of retrieving relevant tweets to a set of news has been pointed out in recent research \cite{kothari2013detecting} with ad-hoc retrieval techniques achieving low Recall (0.5). Prior work relies mostly on tweets where the url of the article is explicitly provided, therefore obtaining a clean but potentially small set of tweets. Our initial tweet-retrieval process gathers a large set of potentially relevant tweets, which we carefully filter in a following step, using a machine learning approach. Since we aim at capturing and continuously tracking active conversations around particular news stories, it is important to not have to artificially restrict the set of tweets from which we extract hashtags. 
 Table \ref{table:auto-stream-stats} shows statistics on the number of tweets in the daily Twitter streams over the 2 months data. The minimum tweet activity happened on Sunday, October 20, 2013, and the maximum activity corresponds to October 30, 2013, an eventful day for  Champions League (european soccer competition). Table \ref{table:raw-tweets} gives a sample of tweets from the tweet-bag of our example article.
\begin{table}
\centering
\caption{Statistics on the daily number of tweets in our focused (dynamic-keyword) Twitter stream.}
\label{table:auto-stream-stats}
\begin{tabular}{|c|c|c|c|c|}
\hline
Min. & Median & Mean & Max & Stdev \\\hline
  91,828  &  350,833 &  424,778 & 1,389,265 & 245,536.3 \\\hline
 \end{tabular}
\end{table}
\begin{table}[t]
\centering
\caption{Example tweets  with hashtag \#seanad, from the tweet-bag of Table \ref{table:example-article} article.}
\label{table:raw-tweets}
\begin{tabular}{|p{8cm}|}
\hline
Like many of his cabinet, the receive knob is broken, so he's permanently on transmit \#enda \#seanad \\ \\
Poor Enda Kenny @MayoGAA \#stillhurting "We know now that like the All Ireland Final, it is not going to be replayed... \#seanad  \\ \\
Jaysus Inda Kenny says no \#seanadref rerun. So, only feckin thing we've 2 look forward 2 is Fine Gael night of d'long knives \#vinb \#seanad \\ \\
So Kenny decides today to reform the Seanad after spending millions on a referendum-what an ejit we have running this country\#waste\#seanad \\ \\
FG fears day of reckoning over Enda Kenny \#Seanad gamble - The Irish Times - Mon, Oct 07, http://t.co/8FtnVxV86d \\\hline
 \end{tabular}
\end{table}
\section{Learning Algorithm for Scoring Hashtags}

In this section we discuss the process of modeling hashtag selection as a learning problem. 
We parse the stream of news articles and the Twitter stream daily, in order to extract and rank hashtags for each news article. For tweet-processing, we remove stop words, punctuation, URLs and user names, and apply stemming to the remaining terms.
For each day, and each news article, we separate the tweets of the corresponding Twitter stream \emph{per article}, based on a shallow matching of tweet keywords and article-keywords (as extracted for the Streaming API and showcased in Table \ref{table:keywords-permutation}). 
This results in a local tweet-bag per article, that can be analyzed for extracting hashtags and hashtag information, e.g., frequency, keyword-profile describing the hashtag as reflected in its tweet-bag.
Next, we form article-hashtag pairs, and compute features of each pair useful for discriminating whether a hashtag is relevant to a given article.

\noindent{\bf Features.} 
Currently, we extract four features for each article-hashtag pair, two features that characterize the local hashtag profile, while the other two characterize the global hashtag profile, useful for describing specific versus general hashtags, as shown in Table \ref{table:features-learning}. 

One of the first features we select is the cosine similarity between the tf.idf keyword profile of the article, and that of the local hashtag profile (as extracted from the tweet-bag). 
To avoid noise in the article profile, rather than selecting terms from the full article-body, we only select them from the headline and sub-headline, but compute their tf.idf weight using the entire article.
Additionally, we extract the local popularity of the hashtag, i.e., the number of tweets in the article tweet-bag, mentioning that hashtag.  

We extract the global frequency of the hashtag in the entire Twitter stream (rather than only the local tweet-bag of the article), and we compute the cosine between the local and the global hashtag keyword-profile, 
to asses how much does the global hashtag profile diverge from the local profile.
Note that globally, the same hashtag may refer to different events, or a hashtag may be preferred over a time window to refer to a certain event, and then 
slowly discarded or outweighed by other hashtags. Therefore, using local and global features for each hashtag, addresses the issue of time-of-use and scope of a hashtag.
For each article-hashtag pair, we now have four features describing how relevant a hashtag may be for a given article. We normalize all four features to the $[0,1]$ interval.
Next, we discuss how to use these features and a set of manually labeled article-hashtag pairs for learning to identify relevant hashtags.
\begin{table*}[t]
\centering
\caption{Features of learning algorithm for scoring hashtags.}
\label{table:features-learning}
\begin{tabular}{|c|p{11cm}|}
\hline
Features & Definition \\\hline
Local Frequency & The number of tweets in the article tweet-bag, mentioning that hashtag.\\\hline
Local Cosine Similarity &  Cosine similarity between the tf.idf keyword profile of the article, and that of the local hashtag profile. \\\hline
Global Frequency & The number of tweets in the entire Twitter stream, mentioning that hashtag. \\\hline
Global Cosine Similarity & Cosine similarity between the local and the global tf.idf hashtag keyword profile (measures how specialized a hashtag is).\\\hline
 \end{tabular}
\end{table*}

\noindent{\bf Labeled Data.}
In order to build a classification algorithm for recognizing relevant hashtags, we need labeled article-hashtag pairs. We selected two days at random from the two month dataset, October 23, and November 23, 2013, extracted all the article hashtag pairs and their features as described above, and asked two annotators to manually label each pair. Table \ref{table:manual-labels-stats} shows statistics on the number of article-hashtag pairs extracted.
\begin{table}[!]
\centering
\caption{Statistics on articles and article-hashtag pairs extracted for the manually labeled dataset.}
\label{table:manual-labels-stats}
\begin{tabular}{|c|c|c|c|c|}
\hline
Date			&	Total articles 	& With hashtags 	& Total pairs \\\hline
Oct23 	& 	146			&	90			&	1,107 \\\hline
Nov23 	& 	142			&	87			&	1,395 \\\hline
 \end{tabular}
\end{table}
The data annotation process is very intensive, since for many pairs it is hard to assess whether the hashtag is relevant or not, without reading the article fully, and searching Twitter for the particular hashtag, or closely analyzing the hashtag-local-profile. In our study, we have found several categories of hashtags, as shown in Table \ref{table:example-article}.
Given the difficulty of this task, we asked the annotators to decide which of the three scenarios applies to each pair: (1) a hashtag is \emph{specific and relevant} to the topic of the news article (e.g., \#seanad, \#seanref, for the example article in Table \ref{table:example-article}), (2) \emph{general and relevant} (e.g., \#rtepm, \#news), or (3) \emph{irrelevant} (e.g., \#tobaccodirective, is related to Enda Kenny, but on a different topic).  For abusive hashtags, the annotators were advised to decide depending on the local context, e.g., \#caughtrotten is relevant to our example article since it groups users discussing the impact of the negative referendum result on the leading party Fine Gael. For the purpose of our experiments, we merged the first two classes into simply relevant (a positive example in binary classification) or irrelevant (negative example).
The inter-annotator agreement was 80\%. We used the subset of examples where both annotators agreed for training/testing a classification algorithm.

\noindent{\bf Classification Algorithm}
We train and test our approach by employing a series of Weka \cite{witten2011data} classification algorithms. 
The algorithm only sees the examples as described by the four features, and can learn thresholding strategies on the provided features. For example, to classify a hashtag as relevant for a given article, a classification algorithm may learn (from the training set) that the cosine feature should be higher than 0.5 and the hashtag frequency should be close to 1. Additionally, most classifiers provide a score describing the likelihood that a hashtag is relevant to the article. We use this classification score to rank hashtags for each article. 

\section{Evaluation}

In order to evaluate our overall strategy for retrieving, learning, and ranking hashtags, we present three evaluation settings. 

\begin{itemize}
\item {\bf Small}: We use the manual labeling of article-hashtag pairs for two random days October 23, 2013 (90 articles with hashtags) for training, and November 23, 2013 (87 articles with hashtags), for testing. 
There are 874 training examples (39.7\% positives, 61.3\% negatives) and 1,122 test examples (45.7\% positives, 54.3\% negatives).

\item {\bf Medium:} Many of the Irish Times articles are discussed on Twitter, by posting the URL (or short URL) of the article directly in the tweet. 
We parse all tweets from our focused Twitter stream, containing URLs of Irish Times articles (we expand the short URL and only retain genuine URLs pointing to real Irish Times articles), and extract the Twitter-user-assigned hashtags for those articles. 
We then use the user-assigned-hashtag data as a form of ground truth, by assuming all user-assigned hashtags are relevant. We train a classifier on the manually labeled data, 
and check how many user-hashtags can our learning algorithm recognize.
There are 1,136 articles with user-hashtags, that lead to 2,773 article-user-hashtag pairs, extracted from 2,801 tweets.
In the intersection between the article-user-hashtag pairs and RSS article-hashtag pairs, there are 732 articles, 
 and 	1,146 articles-user-hashtag test examples.
  Note that although this experiment covers many more articles than the previous one (732 versus 87), 
 the user-hashtag set is smaller than the hashtag set retrieved by our technique, therefore the test data only contains 1,146 pairs.
We train on 2,502 example pairs (October 23, and November 23, 2013) and test on the 1,146 article-user-hashtag examples.

\item {\bf Large:} For each article in our two month set, we attempt to assign hashtags with the procedure described above. 
We train a classifier based on the manually labeled data and the user assigned hashtags (3,648 training examples), and test it on all the article-hashtag pairs retrieved, to assign a classification score to each article-hashtag pair. 
There are 3,388 articles that get assigned hashtags 
and a total of 62,764 article-hashtag example pairs. 
For each article that gets relevant hashtags (i.e., it gets a classification score above 0.5), we rank the hashtags assigned, and take top-3 most relevant hashtags with respect to the classification score. This process results in 2,838 articles with at least one hashtag classified as relevant. Out of this set, we randomly select 422 articles for manual evaluation (about 15\% of all articles with relevant hashtags). This results in 1,029 article-hashtag test pairs to be manually evaluated.
\end{itemize}

\subsection{Error Metrics}
We employ metrics from both machine learning and information retrieval to assess the quality of our results.

\noindent{\bf Classification.}
We compute the following standard binary classification metrics \cite{witten2011data}, where $TP$ stands for true positive, $FP$ for false positive, $TN$ true negative and $FN$ false negative:
\begin{small}
$$Accuracy = \frac{TP+TN}{TP+FP+TN+FN}$$
$$Precision=\frac{TP}{TP + FP} ;\; Recall=\frac{TP}{TP+FN}$$
$$F1=\frac{2 \cdot Precision \cdot Recall}{Precision+Recall}$$ 
\end{small}
The above metrics assume a fixed classification threshold, but one can typically vary this threshold to improve classifier performance (e.g., by tuning on the training set). Rather than focusing on a binary split into positives and negatives, other metrics characterise the ranking performance of a classifier.
The ROC curve characterises the performance of a binary classifier, by capturing the fraction of true positives versus that of false positives at varying classification thresholds \cite{fawcett2004roc}.
 The area under the ROC curve ($AUC$) is an aggregate measure corresponding to the probability that a randomly chosen positive instance is ranked higher than a randomly chosen negative instance.  An $AUC$ of 1 represents perfect ranking of all positives above all negatives.
 
 \noindent{\bf Information Retrieval.}
 The above metrics assess the classification algorithm over all article-hashtag pairs. In order to assess the per article hashtag ranking quality, we also employ metrics from information retrieval.
 We evaluate the classifier-induced ranking of hashtags, for each article,
by the Precision@1 and the Normalized Discounted Cumulative Gain (NDCG) \cite{jarvelin2000ir,DBLP:conf/colt/WangWLHL13} as defined below.
$$DCG@k = \sum_{i=1}^{k}\frac{2^{rel_i} - 1}{log_2(i+1)}$$
$$NDCG@k = \frac{DCG@k}{IdealDCG@k}$$
These are standard information retrieval metrics for evaluating the quality of a ranking function \cite{manning2008introduction}. The Precision@1 captures how satisfied the user is with the best ranked hashtag for each article. It is computed as the number of times that a relevant hashtag is in the first position of the ranking, weighted by the relevance score and normalized.
The NDCG describes the cumulative gain the user obtains by examining the retrieval results up to a given rank position $k$.
NDCG makes it possible to evaluate the ranking of hashtags uniformly across articles (independent of whether the article gets $k$ or less relevant hashtags). 
It further allows a more fine-grained evaluation, by penalising more the relevant results placed at low ranks.
 
\subsection{Results: Small Experiment}

For the first experimental setting, we had 874 training examples (October 23, 2013) and 1,122 test examples (November 23, 2013) on which the labels of both annotators agreed. 

\noindent{\bf Baselines.}
In order to assess the actual utility of a learning approach, we first evaluate two simple baseline techniques. On the test set (November 23, 2013), we select the top-3 hashtags per article (257 pairs out of 1,122), using the highest local hashtag frequency and the highest local cosine similarity. 
Table \ref{table:baselines-small} shows the precision of results using these two simple heuristics.
\begin{table}[t!]
\centering
\caption{{\bf Small} (baselines): Top-3 relevant hashtags using most frequent or highest cosine.}
\label{table:baselines-small}
\begin{tabular}{|l|c|}
\hline
Baselines				& Precision 	\\\hline
Most Frequent Top-3 	& 0.548	\\
Highest Cosine Top-3 	& 0.634	 \\\hline
\end{tabular}
\end{table}

\noindent{\bf Learning Approach.}
We now evaluate the classifier's ability to retrieve all the hashtags deemed relevant by our annotators as well as its ability to rank them before the irrelevant ones. 
We experimented with a series of Weka classifiers, with default parameter settings. MultilayerPerceptron, Logistic (regularised logistic regression) and Kstar (K-nearest neighbours with entropy-based-distance) delivered the best results, as shown in Table \ref{table:results-small}.
We note that all three classifiers have high precision (0.85), recall (0.80) and AUC (0.92), showing that the classifier ranks relevant hashtags before irrelevant ones. The AUC is particularly important, since ultimately it is useful to rank the hashtags of each document, from most relevant to least relevant. We also experimented with swapping training and test (i.e., training on November 23, and testing on October 23), and the results are very similar. We note that the Logistic classifier had the highest Precision and AUC. Additionally, the Logistic classifier is a linear model that can be easily interpreted and its classification scores are true probabilities. The Logistic model deemed all four features as important (non-zero weights), with the local cosine feature getting the highest weight, followed by the frequency based features, and ending with the global cosine.
\begin{table*}[htb!]
\centering
\caption{{\bf Small} (learning): Hashtag classification on manually labeled test set.}
\label{table:results-small}
\begin{tabular}{|c|c|c|c|c|c|}
\hline
Weka Classifier				&	Accuracy 		& Precision 	& Recall 	& F1		& AUC \\\hline
MultilayerPerceptron 		& 	84.6\%		& 0.850		& 0.807	& 0.846	& 0.921  \\
Logistic 					& 	84.4\%		& 0.876		& 0.770  	& 0.844	& 0.924	 \\
 Kstar					&	83.9\%		& 0.861     	& 0.774 	& 0.839 	& 0.911 \\\hline
 \end{tabular}
\end{table*}

\subsection{Results: Medium Experiment}

In this setting, we train classifiers on the manually labeled data from the {\bf Small} setting, and test them on Twitter-user-hashtagged data. As training data we analyze three settings, October 23, 2013, November 23, or both days together as training, and article-user-hashtag pairs as test (1,146 test examples). Note that we assume that all the user-assigned hashtags are relevant, which may not necessarily be the case, since sometimes users also assign spurious hashtags, e.g., \#annoying \#omg. Our algorithm may consider such hashtags as irrelevant, in particular if the global hashtag profile strongly diverges from the local hashtag profile. Since we assume all the test examples are relevant (there are no negative test examples), in this setting the Accuracy is the same as Recall.
We again test the two baselines to check Recall at top-3, as shown in Table \ref{table:baselines-medium}.  In Table \ref{table:results-medium} we show Recall over all Twitter-user-hashtags retrieved by our algorithm as being relevant (classification score above 0.5).
We note that when training only on October or November 23, the classifier retrieves about 80\% of user-hashtags as relevant, while when we increase the amount of training data, by combining the October and November examples,  the accuracy of MultilayerPerceptron stands at 84.5\%, a similar value to that of the small setting experiment.
\begin{table}[t!]
\centering
\caption{{\bf Medium} (baselines): Top-3 relevant hashtags using most frequent or highest cosine.}
\label{table:baselines-medium}
\begin{tabular}{|l|c|}
\hline
Baselines				& Recall 	\\\hline
Most Frequent Top-3 	& 	0.503		\\
Highest Cosine Top-3 	& 	0.644		\\\hline
\end{tabular}
\end{table}
\begin{table}[t!]
\centering
\caption{{\bf Medium} (learning): Hashtags identified as relevant on the Twitter-user-labeled dataset.}
\label{table:results-medium}
\begin{tabular}{|c|c|c|}
\hline
Training Data		& Weka CLassifier		&	Recall \\\hline
Oct23			& MultilayerPerceptron 	& 	0.781	  \\
				& Logistic 				& 	0.756	 \\
 				& Kstar				&	0.704	 \\\hline
Nov23			& MultilayerPerceptron 	& 	0.792	  \\
				& Logistic 				& 	0.808	 \\
 				& Kstar				&	0.787	 \\\hline
Oct23 \& Nov23	& MultilayerPerceptron 	& 	0.845	  \\
				& Logistic 				& 	0.797	 \\
 				& Kstar				&	0.776	 \\\hline
\end{tabular}
\end{table}

\subsection{Results: Large Experiment}

Building on observations from the previous two evaluation settings, we train and apply a classifier to the entire two month data collection. For training we use the labeled examples of October 23, November 23, and the Twitter-user-hashtagged examples. We then apply this classifier to all the article-hashtag pairs extracted from the RSS and Twitter streams, a total of 62,764 test examples (3,388 unique articles).
We extract only articles that get at least one relevant hashtag (based on classification score above 0.5; 2838 articles), and manually asses a random sample (422 articles, 1,029 pairs), using 0 and 1 relevancy scores. We evaluate both the filtering quality, i.e., the classification across all article-hashtag pairs (to asses the Precision over the pairs classified as relevant), as well as the hashtag ranking quality per article, using information retrieval metrics. For the article oriented metrics, we use Precision@1 and NDCG@3 and average them across all articles. 
We show the average values and the t-test 95\% confidence interval for Precision@1 and NDCG@3 across all articles in Table \ref{table:results-large}.
\begin{table*}[htb!]
\centering
\caption{{\bf Large} (learning): Hashtag filtering and ranking on two month article-dataset.}
\label{table:results-large}
\begin{tabular}{|l||l|l|}
\hline
Precision \small{(over 1,029 pairs)}		&	Precision@1 \small{(over 422 articles)}	& NDCG@3 \small{(over 422 articles)} \\\hline
0.869							&	0.900 \small{([0.871, 0.929], $p<2.2e-16$)}			& 0.877 \small{([0.850, 0.904], $p<2.2e-16$)} \\\hline
 \end{tabular}
\end{table*}
We note that the precision for the filtering step (binary classification into relevant/irrelevant) is fairly high (Precision 0.86), and similar to what we have seen in the previous experiments. When we evaluate the quality of ranking of hashtags for each article, we see a similar result: the Precision@1 is 0.9, while the NDCG@3 which penalizes relevant hashtags ranked at low ranks, is 0.87.

\subsection{Discussion}

In order to make the whole methodology more explicit, in Table \ref{table:discussion-results} we show some example articles from our annotated sample of the {\bf Large} setting, their extracted keywords,
their (up to top-3) ranked hashtags, together with the features extracted for the corresponding pair, the classifier score and the annotator relevance score.
We observe that for local as well as international news (first 3 articles), the hashtags assigned and ranked (by classifier score) are relevant and quite specific (e.g., \#ecb, \#walshwhiskeydistillery). 

We found three main reasons why an article does not get any (relevant) hashtag: the article-keyword extraction process is faulty (due to part-of-speech tagging errors or due to the fact that the extracted keywords are too generic); there is no discussion on Twitter about that particular news story; the tweets relevant to an article do not contain any hashtags.
The aspect of assigning noisy or irrelevant hashtags can be mitigated to some extent by tuning the classifier threshold (here we used the default classification score of 0.5).
Additionally, the four features describing each article-hashtag pair could be enhanced, e.g., using user authority to re-weight tweets, filtering spammy hashtags (e.g., \#ff, \#followback).
Regarding the lack of hashtags in the tweet-bag of an article, in such cases we could employ recent techniques for extracting informative tweets \cite{vstajner2013automatic}, or adapt our approach for the problem of assigning Twitter users (rather than hashtags) relevant to a given news article.
Among the categories of news articles from our dataset that get a lot of relevant hashtags, are those related to sporting events. This is likely due to the fact that sporting events get a lot of reaction on Twitter.  
The manual annotation for the learning approach is potentially noisy, since at times it is quite difficult to decide whether a hashtag is relevant or not, without considerable background knowledge. In this respect we plan to employ crowd sourcing platforms such as Crowdflower, in order to obtain larger and possibly cleaner labeled datasets.
\begin{table*}[t!]
\centering
\begin{small}
\caption{Example results from our two-month annotated sample.}
\label{table:discussion-results}
\begin{tabular}{|l|l|l|l|l|l|l|c|c|}
\hline
Article URL 										& Article Keywords		& Hashtag				&  LFr 	& LCo	& GFr 	& GCo 	& ClassifScor 	& RelScor \\\hline
tech-titans-in-town-for-dublin						& dublin, dubstarts,		& \#websummit				& 1.00 	& 0.35	& 0.58	& 0.82	& 0.92		& 2 \\	
-web-summit-1.1573638							& summit, tech, web		& \#tech					& 0.23	& 0.45	& 0.70	& 0.37	& 0.89		& 1 \\	
												&						& \#web					& 0.42	&0.41	& 0.40	& 0.52	& 0.82		& 2 \\\hline
whiskey-distillery-to-create-						& carlow, co, distillery,		&\#whiskey				& 1.0	0	& 0.73	& 0.16	& 0.56	& 0.99		& 2 \\				
55-jobs-for-co-carlow-1.1562843					& walsh, whiskey			&\#carlow					& 0.90	& 0.64	& 0.16	& 0.53	& 0.99		& 2 \\
												& 						& \#walshwhiskeydistillery	& 0.66	& 0.61	& 0.00	& 1.00	& 0.97		&2 \\\hline
ecb-s-draghi-moves-to-ease-fears-					& banks, draghi,ecb		& \#ecb					& 1.00	& 0.50	& 0.39	& 0.42	& 0.97		&2 \\	
on-interest-rates-1.1604077						&						& \#draghi					& 0.54	& 0.58	& 0.19	& 0.69	& 0.96		& 2	\\
												&						& \#news					& 0.00	& 0.46	& 0.89	& 0.29	& 0.90		& 1\\\hline	
climate-change-watchdog-must-be-				& advisory, climate,		& \#delleir					& 1.00	& 0.27	& 0.00	& 1.00	& 0.60		& 0 \\
robust-and-independent-							& council, expert, fiscal	& \#job					& 0.00	& 0.30	& 0.80	& 0.35	& 0.53		& 0 \\
says-report-1.1601809							&					& \#delcfe					& 0.89	& 0.26	& 0.00	& 1.00	& 0.51		& 0 \\\hline
europe-bank-payouts-capped-as-					& capital-europe		& \#europe 				& 0.79	& 0.35	& 0.55	& 0.02	& 0.84	& 1 \\
capital-bar-keeps-rising-1.1603937					& 					& \#travel					& 0.66	& 0.27	& 0.68	& 0.38	& 0.72	& 0 \\\hline
\end{tabular}
\end{small}
\end{table*}

As a toy example show-casing the potential applications of connecting articles and hashtags, in Table \ref{table:articles-of-seanad} we cluster news articles in hashtag space. In this particular example, 
we simply sort news articles that get the hashtag \#seanad, by classifier score. We note that news articles belonging to the same news topic (Seanad referendum) group together, although some do not have any terms in common in the original article headlines. 
\begin{table}[t]
\centering
\caption{Six example articles clustered by hashtag \#seanad based on learning algorithm score.}
\label{table:articles-of-seanad}
\begin{tabular}{|p{5cm}|c|}
\hline
Headline & Classif Score \\\hline
Kenny embroiled in tense spat over referendum at FG meeting. & 0.99 \\
The political reform shortfall remains. & 0.97 \\
Taoiseach anxious to see Seanad become effective watchdog. & 0.92 \\
FG fears day of reckoning over Enda Kenny's Seanad gamble. & 0.84 \\
Government says Seanad reform is on the agenda. & 0.83 \\
Final arguments made in appeal against Callely decision. & 0.82 \\\hline
\end{tabular}
\end{table}

\section{Conclusion}

In this work we present a framework for connecting news articles to their relevant Twitter conversations, as semantically grouped by Twitter hashtags.
We discuss the aspect of continuously tracking a stream of news and tweets, and present an approach for obtaining a large focused Twitter stream automatically 
seeded by a dynamic keyword set extracted from the articles. 
Furthermore, we model the problem of hashtag assignment as a classification problem, and analyze a framework for hashtag retrieval and appropriate features and data 
for building a hashtag classifier. We evaluate our methods and show that our approach achieves high precision for this task.
%

\noindent{\bf Future Work.} We plan to extend our study to track several RSS news feeds and Twitter conversations, and test a prototype with journalists.
We also intend to investigate applications of our methods to clustering of articles in hashtag space, story tracking and event detection.
%



\end{document}